\documentclass[runningheads]{llncs}
\usepackage{ifthen}
\usepackage[pdftex]{graphicx}
\graphicspath{{figs/}{./}}
\usepackage{calc}
\begin{document}

\title{PlDoc: Wiki style Literate Programming for Prolog}
\titlerunning{PlDoc}

\author{Jan Wielemaker\inst{1} and Anjo Anjewierden\inst{2}}
\authorrunning{Wielemaker and Anjewierden}
\institute{Human-Computer Studies Laboratory,
	   University of Amsterdam, \\
	   Kruislaan 419,
	   1098 VA Amsterdam,
	   The Netherlands, \\
	   \email{wielemak@science.uva.nl}
	   \and
	   Department of Instructional Technology, 
	   Faculty of Behavourial Sciences,
	   University of Twente, \\
	   PO Box 217, 7500 AE Enschede, The Netherlands \\
	   \email{a.a.anjewierden@utwente.nl}}

\maketitle
\bgroup
\input myllncs.sty
\setcounter{topnumber}{3}
\setcounter{bottomnumber}{1}
\setcounter{totalnumber}{3}
\renewcommand{\topfraction}{.9}
\renewcommand{\bottomfraction}{.9}
\renewcommand{\textfraction}{.1}
\renewcommand{\floatpagefraction}{.9}

\begin{abstract}
This document introduces PlDoc, a literate programming system for
Prolog. Starting point for PlDoc was minimal distraction from the
programming task and maximal immediate reward, attempting to seduce the
programmer to use the system. Minimal distraction is achieved using
structured comments that are as closely as possible related to common
Prolog documentation practices. Immediate reward is provided by a
web interface powered from the Prolog development environment that
integrates searching and browsing application and system documentation.
When accessed from \jargon{localhost}, it is possible to go from
documentation shown in a browser to the source code displayed in
the user's editor of choice.
\end{abstract}


\section{Introduction}

Combining source and documentation in the same file, generally named
\jargon{literate programming}, is an old idea. Classical examples are
the \TeX{} source \cite{479} and the self documenting editor GNU-Emacs
\cite{806466}. Where the aim of the \TeX{} source is first of all
documenting the program, for GNU-Emacs the aim is to support primarily
the end user. A more recent success story is
JavaDoc\footnote{\url{http://java.sun.com/j2se/javadoc/}}.

There is an overwhelming amount of articles on literate programming,
most of which describe an implementation or qualitative experience using
a literate programming system \cite{ramsey91literate}. Shum and Cook
\cite{191059} describe a controlled experiment on the effect of literate
programming in education. Using literate programming produces more
comments in general. More convincingly, it produced `how documentation'
and examples where, without literate programming, no examples were
produced at all. Nevertheless, subjects using literate programming (in
this case AOPS, \cite{AOPS}) was considered confusing and harmed
debugging the program.

Recent developments in programming environments and methodologies make a
case for re-introducing literate programming \cite{1035054}. The success
of systems such as Doxygen \cite{doxygen} based on some form of
structured comments in the source code, making the literate programming
document a valid document for the programming language is evident. Using
a source document that is valid for the programming language ensures
smooth integration with tools designed for the language.

Note that these developments are different from what Knuth intended:
``The literate programmer can be regarded as an essayist that explains
the solution to a human by crisply defining the components and
delicately weaving them together into a complete artistic creation''
\cite{479}. Embedding documentation source code comments merely produces
an \jargon{API Reference Manual}.

In the Prolog world we see lpdoc \cite{DBLP:conf/cl/Hermenegildo00},
documentation support in the Logtalk \cite{pmoura03} language and the
ECLiPSe Document Generation
Tools\footnote{\url{http://eclipse.crosscoreop.com/doc/userman/umsroot088.html}}
system. All these approaches use Prolog \jargon{directives} making
additional statements about the code that feed the documentation system.
In 2006 a commercial user in the UK whose products are developed using a
range of technologies (including C++ using Doxygen for documentation)
approached us to come up with an alternative literate programming system
for Prolog, aiming at a documentation system as non-intrusive as
possible to their programmers' current practice.

This document is structured as follows. First we outline the different
options available to a literate programming environment and motivate our
choices. Next we introduce PlDoc using and example, followed by a more
detailed overview of the system. \Secref{toko} tells the story of
introducing PlDoc in a large open source program, while we compare our
work to related projects in \secref{related}.

\postscriptfig[width=\linewidth]{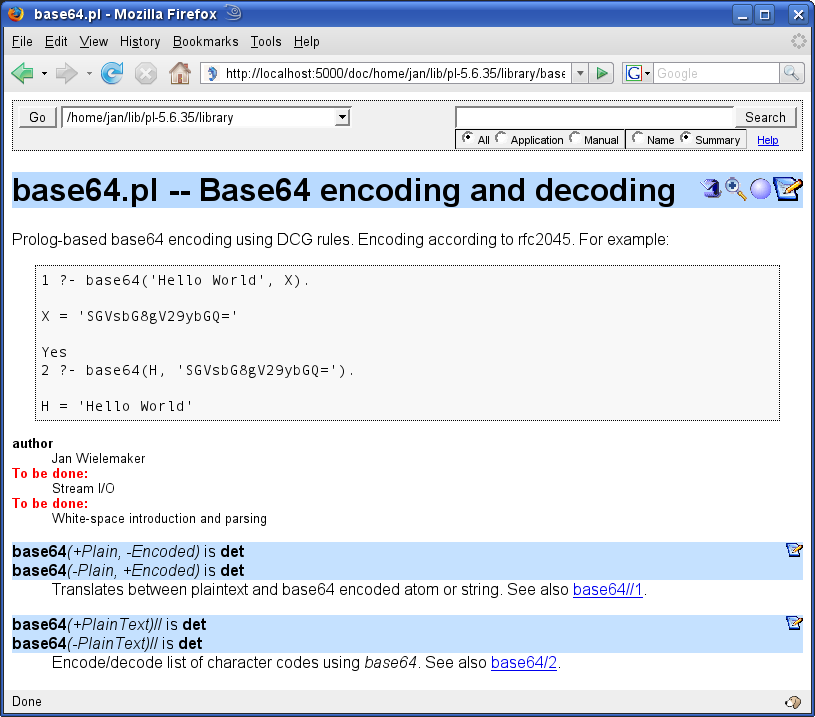}{Documentation of library
base64.pl.  Accessed from `localhost', PlDoc provides edit and reload
buttons.}

\section{An attractive literate programming environment}

Most programmers do not like documenting code and Prolog programmers are
definitely no exception to this rule. Most can only be `persuaded' by
the organisation they work for, using a documentation biased grading
system in education \cite{191059} or by the desire to produce code that
is accepted in the Open Source community. In our view we must seduce the
programmer to produce API documentation and internal documentation by
creating a rewarding environment. In this section we present the
available choicepoints and motivate our primary choices from these
starting points.

For the design of a literate programming system we must make decisions
on the input: the language in which we write the documentation and how
this language is merged with the programming language (Prolog) into a
single source file. Traditionally the documentation language was \TeX{}
based (including Texinfo). Recent systems (e.g.\ JavaDoc) also use HTML.
In Prolog, we have two options for merging documentation in the Prolog
text such that the combined text is a valid Prolog document. The first
is using Prolog comments and the second is to write the documentation in
\jargon{directives} and define (possibly dummy) predicates that handle
these directives.

In addition we have to make a decision on the output format. In backend
systems we see a shift from \TeX{} (paper) and plain-text (online)
formats towards HTML, XML+XSLT and (X)HTML+CSS which are widely
supported in todays development environments.  Web documents provide
both comfortable online browsing and reasonable quality printing.

\postscriptfig[width=\linewidth]{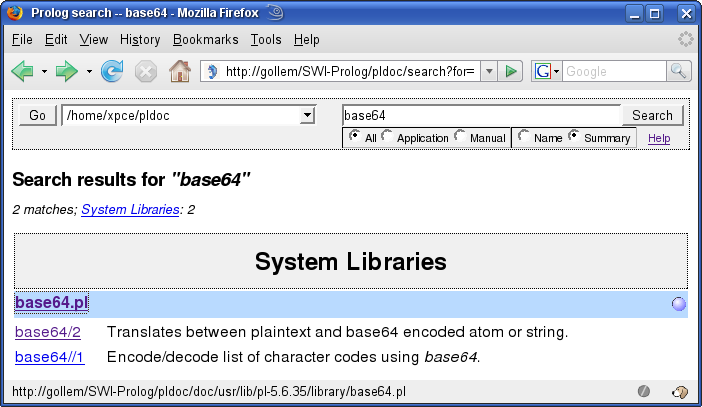}{Searching for ``base64''}

In this search space we aim at a system with little overhead for the
programer and a short learning curve that immediately rewards the
programmer with a better overview and integrated web-based search over
both the application documentation and the Prolog manual.

\paragraph{Minimal impact}

Minimising the impact on the task of the programmer is very important.
Programming itself is a demanding task and it is important to reduce the
mental load to the minimum, only keeping that what is essential for the
result. Whereas object oriented languages can extract some basics from
the class hierarchy and type system, there is little API information
that can be extracted automatically from a Prolog program, especially if
it does not use modules. Most information for an API reference must be
provided explicitly and additionally to the program.

Minimising impact as well as maximising portability made us decide
against the approach of lpdoc, ECLiPSe and Logtalk which provide the
documentation in language extensions by means of directives and in
favour of using structured comments based on layout and structuring
conventions around in the Prolog community. Structured comments start
with \verb$%%$ (similar to PostScript document structuring comments) and
use Wiki \cite{wikiway} structuring conventions extended with Prolog
conventions such as referencing a predicate using \bnfmeta{name}/\bnfmeta{arity}. Wiki
is a simple plain-text format designed for collaborative interactive
management of web pages. Wikis differ in the details on the text format.
We are particularly interested in wiki formats based on common practice
simulating font and structuring conventions in plain text such as
traditional email, usenet and comments in source code.

\paragraph{Maximal reward to the programmer}

A system is more easily accepted if it not only provides reward for the
users of the software module, but also to the programmer him/herself. We
achieve this by merging the documentation of the loaded Prolog code with
the Prolog manuals in a consistent view presented from a web server
embedded in the development environment. This relieves the programmer
from making separate searches in the manuals and other parts of system
under development.

\paragraph{Immediate reward to the programmer}

Humans love to be rewarded immediately. This implies the results must be
accessible directly. This has been achieved by adding the documentation
system as an optional library to the Prolog development environment.
With PlDoc loaded into Prolog, the compiler processes the structured
comments, maintaining a Prolog database as described in
\secref{processdoc}. This database is made available to the developer
through a web server running in a separate thread (\secref{publish}).
The SWI-Prolog \index{make/0}\predref{make}{0} comment updates the running Prolog system to the
latest version of the loaded sources and updates the web site at the
same time.

\section{An example}
\label{sec:example}

Before going into detail we show the documentation process and access
for the SWI-Prolog library \file{base64.pl}, providing a DCG rule for
base64 encoding and decoding as well as a conversion predicate for
atoms. Part of the library code relevant for the documentation is in
\figref{base64code}. We see a number of special constructs:

\begin{figure}
\small
\begin{verbatim}
/** <module> Base64 encoding and decoding

Prolog-based base64 encoding using  DCG   rules.  Encoding  according to
rfc2045. For example:

==
1 ?- base64('Hello World', X).
X = 'SGVsbG8gV29ybGQ='

2 ?- base64(H, 'SGVsbG8gV29ybGQ=').
H = 'Hello World'
==

@tbd    Stream I/O
@tbd    White-space introduction and parsing
@author Jan Wielemaker
*/

%%      base64(+Plain, -Encoded) is det.
%%      base64(-Plain, +Encoded) is det.
%
%       Translates between plaintext and base64  encoded atom or string.
%       See also base64//1.

base64(Plain, Encoded) :- ...

%%      base64(+PlainText)// is det.
%%      base64(-PlainText)// is det.
%
%       Encode/decode list of character codes using _base64_.  See also
%       base64/2.

base64(Input) --> ...
\end{verbatim}

\noindent
    \caption{Commented source code of library base64.pl}
    \label{fig:base64code}
\end{figure}

\begin{itemize}
    \item The \verb$/** <$\verb$module> Title$ 
comment introduces overall documentation of the module. Inside, the
\verb$==$ delimited lines start a source code block. The
\verb$@$\arg{keyword} section provides JavaDoc inspired keywords from a
fixed and well defined set (see end of \secref{syntax}).

    \item The \verb$%%$ comments start with one or more \verb$%%$ lines that
contain the predicate name, argument names with optional mode, type
and determinism information. Multiple modes and predicates can be
covered by the same comment block. This is followed by wiki text,
processed using the same rules that apply to the module comment.
Like JavaDoc, the first sentence of the comment body is considered
a \jargon{summary}. Keyword search processes both the formal description
and the summary. Keyword search on format aspects and a summary line
have a long history, for example in the Unix \program{man} command.
\end{itemize}

\section{Description of PlDoc}
\label{sec:overview}

\subsection{The PlDoc syntax}
\label{sec:syntax}

PlDoc processes structured comments.  Structured comments are Prolog
comments starting with \verb$%%$ or \verb$/**$.  The former is more 
in line with the Prolog tradition for commenting predicates while the
second is typically used for commenting the overall module structure.
The system does not enforce this.  Java programmers may prefer using
the second form for predicate comments as well.

Comments consist of a formal header, a wiki body and JavaDoc inspired
keywords. When using \verb$%%$ style comments, the formal header ends
with the first line with a single \verb$%$. Using \verb$/**$ style
comments the header is ended by a blank line. The header is either
``\bnfmeta{module} \arg{Title}'' or one or more predicate head declarations.
The \bnfmeta{module} syntax can be extended easily.

The type and mode declaration header consists of one or more Prolog
terms.  Each term describes a mode of a predicate.  The syntax is
described in \figref{headbnf}.

\begin{figure}
\begin{center}
\begin{tabular}{lrl}
\hline
\bnfmeta{modedef}	\isa \bnfmeta{head}['//'] ['is' \bnfmeta{determinism}] \\
\bnfmeta{determinism}	\isa 'det' \\
		\ora 'semidet' \\
		\ora 'nondet' \\
		\ora 'multi' \\
\bnfmeta{head}		\isa \bnfmeta{functor}'('\bnfmeta{argspec} \{',' \bnfmeta{argspec}\}')' \\
		\ora \bnfmeta{atom} \\
\bnfmeta{argspec}	\isa [\bnfmeta{mode}]\bnfmeta{argname}[':'\bnfmeta{type}] \\
\bnfmeta{mode}	        \isa '+' $\mid$ '-' $\mid$ '?' $\mid$ ':'
		     $\mid$ '@' $\mid$ '!' \\
\bnfmeta{type}		\isa \bnfmeta{term} \\
\hline
\end{tabular}
\end{center}
    \caption{BNF for predicate header}
    \label{fig:headbnf}
\end{figure}

The optional \verb$//$-postfix indicate \bnfmeta{head} is a DCG rule. The
\jargon{determinism} values originate from Mercury
\cite{DBLP:conf/acsc/JefferyHS00}. Predicates marked as \const{det} must
succeed exactly once and not leave any choice points. The
\const{semidet} indicator is used for predicates that either fail or
succeed deterministically. The \const{nondet} indicator is the most
general one and implies there are no constraints on the number of times
the predicate succeeds and whether or not it leaves choice points on the
last success. Finally, \const{multi} is as \const{nondet}, but demands
the predicate to succeed at least one time. Informally, \const{det} is
used for deterministic transformations (e.g.\ arithmetic),
\const{semidet} for tests, \const{nondet} and \const{multi} for
\jargon{generators}.

The mode patterns are given in \figref{modes}. Originating from DEC-10
Prolog were the \jargon{mode} indicators (\verb$+$,\verb$-$,\verb$?$)
had a formal meaning. The ISO standard \cite{Deransart:96} adds
`\verb$@$', meaning ``the argument shall remain unaltered''. Quintus
added `\verb$:$', meaning the argument is module sensitive. Richard
O'Keefe
proposes\footnote{\url{http://gollem.science.uva.nl/SWI-Prolog/mailinglist/archive/2006/q1/0267.html}}
`\verb$=$' for ``remains unaltered'' and adds `\verb$*$' (ground) and
`\verb$>$' ``thought of as output but might be nonvar''.

\begin{figure}
\begin{center}
\begin{tabular}{cp{0.8\linewidth}}
\hline
+ &	Argument must be fully instantiated to a term that satisfies the
	type. \\
- &	Argument must be unbound on entry. \\
? &	Argument must be bound to a \emph{partial term} of the indicated
	type.  Note that a variable is a partial term for any type. \\
: &	Argument is a meta argument.  Implies \const{+}. \\
@ &	Argument is not further instantiated. \\
! &	Argument contains a mutable structure that may be modified using
	\index{setarg/3}\predref{setarg}{3} or \index{nb_setarg/3}\predref{nb_setarg}{3}. \\
\hline
\end{tabular}
\end{center}
    \caption{Defined modes}
    \label{fig:modes}
\end{figure}

The body of a description is given to a Prolog defined wiki parser based
on Twiki\footnote{\url{http://www.twiki.org}} using extensions from the
Prolog community.  In addition we made the following changes.

\begin{itemize}
    \item List indentation is not fixed, the only requirement is that
    all items are indented to the same column.

    \item Font changing commands such as \verb$*bold*$ only work if
    the content is a single word. In other cases we demand \verb$*|bold text|*$.
    This proved necessary due to frequent use of punctuation characters
    in comments that make single font switching punctuation
    characters too ambiguous.

    \item We added \verb$==$ around code blocks (see \figref{base64code}) as
    such blocks are frequent and not easily supported by Twiki.

    \item We added automatic links for \bnfmeta{name}/\bnfmeta{arity}, \bnfmeta{name}//\bnfmeta{arity},
    \bnfmeta{file}.pl, \bnfmeta{file}.txt (interpreted as wiki text) and image
    files using image extensions.  Using \verb$[[$file.png\verb$]]$,
    inline images can be produced.

    \item Capitalised words appearing in the running text that match
    exactly one of the arguments are typeset in \textit{italics}.

    \item We do not process embedded HTML.  One of the reasons is that
    we want the option for other target languages.  Opening up the path
    to unlimited use of HTML complicates this. In addition, passing
    \verb$<$, \verb$>$ and \verb$&$ unmodified to the target HTML easily
    produces invalid HTML.
\end{itemize}

The `@' keyword section of a comment block is heavily based on JavaDoc.
We give a summary of the changes and additions below.

\begin{itemize}
    \item \index{@return}\objectname{return} is dropped for obvious reasons.
    \item \index{@error}\objectname{error} is added as a shorthand for \index{@throws}\objectname{throws} error(Error, Context)
    \item \index{@since}\objectname{since} and \index{@serial}\objectname{serial} are not (yet) supported
    \item \index{@compat}\objectname{compat} is added to describe compatibility of libraries
    \item \index{@copyright}\objectname{copyright} and \index{@license}\objectname{license} are added
    \item \index{@bug}\objectname{bug} and \index{@tbd}\objectname{tbd} are added for issue tracking
\end{itemize}

A full definition of the Wiki notation and keywords is in the PlDoc
manual.\footnote{\url{http://www.swi-prolog.org/packages/pldoc.html}}.

\subsection{Processing the comments}
\label{sec:processdoc}

We claimed immediate reward as an important asset. This implies the
documentation must be an integral part of the development environment.
SWI-Prolog aims at providing IDE modules while allowing the user to use
an editor or IDE of choice. An obvious choice is to make the compiler
collect comments and present these to the user through a web interface.
This is achieved using a hook in the compiler called as:
\begin{quote}
\term{prolog:comment_hook}{+Comments, +TermPos, +Term}.
\end{quote}
Here, \arg{Comments} is a list of \arg{Pos}-\arg{Comment} terms
representing comments encountered from where \index{read_term/3}\predref{read_term}{3} started reading
upto the end of \arg{Term} that started at \arg{TermPos}. The calling
pattern allows for processing any comment and distinguishes comments
outside Prolog terms from comments inside the term.

The hook installed by the documentation server extracts structured
comments by checking for \verb$%%$ or \verb$/**$. For structured
comments it extracts the formal comment header and the first line of the
comment body which serves, like JavaDoc, as a \jargon{summary}. The
formal part is processed and the entire structured comment is stored
unparsed, but is associated with the parsed formal header and summary
which are used for linking the comment with a predicate as
well as keyword search. The stored information is available through the
public Prolog API of PlDoc and can be used, together with the cross
referencer Prolog API, as the basis for additional development tools.

\subsection{Publishing the documentation}
\label{sec:publish}

PlDoc realises a web application using the SWI-Prolog HTTP
infrastructure \cite{TPLP06}. Running in a separate thread, the normal
interactive Prolog toplevel is not affected. The documentation server
can also be used from an embedded Prolog system. By default access is
granted to `localhost' only. Using additional options to
\term{doc_server}{+Port, +Options}, access can be granted to a wider
public. A scenario for exploiting this is to have a central Prolog
process with all resources available to a team loaded. Regularly running
update from a central repository and \index{make/0}\predref{make}{0} inside Prolog, it can serve
as an up-to-date and searchable central documentation source. Since
September 15 2006, we host such a server running the latest SWI-Prolog
release with all standard libraries and packages loaded from
\url{http://gollem.science.uva.nl/SWI-Prolog/pldoc/}. Currently
(June 2007), the server handles approximately 100 search requests (1,000
page views) per day.

\postscriptfig[width=\linewidth]{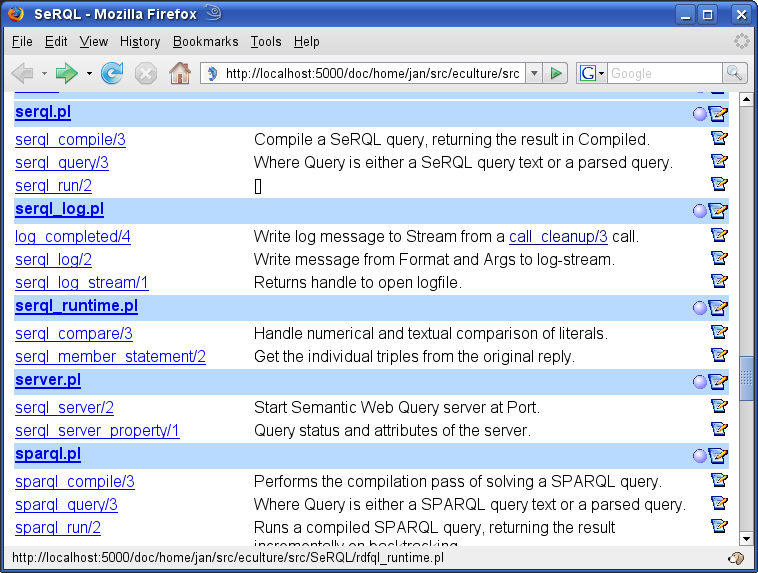}{PlDoc displaying a
directory index with files and their public predicates accessed
from `localhost'.  Each predicate has an `edit' button and each
file a pretty print button (blue circle, see \secref{prettyprint})}

\subsection{IDE integration and documentation maintenance cycle}
\label{sec:maintenance}

When accessed from `localhost', PlDoc by default provides an option to
edit a documented predicate. Clicking this option activates an HTTP
request through Javascript similar to AJAX \cite{ajax}, calling
\term{edit}{+PredicateIndicator} on the development system. This
hookable predicate locates the predicate and runs the user's editor of
choice on the given location. In addition the browser interface shows a
`reload' button to run \index{make/0}\predref{make}{0} and refreshes the current page, reflecting
the edited documentation.

Initially, PlDoc is targeted to the working directory. In the directory
view it displays the README file (if any) and all Prolog files with a
summary listing of the public predicates as illustrated in
\figref{serqlindex}.

As a simple quality control measure PlDoc lists predicates that are
exported from a module but not documented in red at the bottom of
the page.  See \figref{undoc}.

We used the above to provide elementary support through PlDoc for most
of the SWI-Prolog library and package sources (approx.\ 80,000 lines).
First we used a simple \program{sed} script to change the first line of
a \verb$%$ comment that comments a predicate to use the \verb$%%$
notation. Next we fixed syntax errors in the formal part of the
documentation header. Some of these where caused by comments that should
not have been turned into structured comments. PlDoc's enforcement that
argument names are proper variable names and types are proper Prolog
terms formed the biggest source of errors. Finally, directory indices
and part of the individual files were reviewed, documentation was
completed and fixed at some points. The enterprise is certainly not
complete, but an effort of three days made a big difference in the
accessibility of the libraries.

\subsection{Presentation options}
\label{sec:prettyprint}

By default, PlDoc only shows public predicates when displaying a file or
directory index. This can be changed using the `zoom' button displayed
with every page. Showing documentation on internal predicates proves
helpful for better understanding of a module and helps finding
opportunities for code reuse. Searching shows hits from both public and
private predicates, where private predicates are presented in grey using
a yellowish background.

Every file entry has a `source' button that shows the source file.
Structured comments are converted into HTML using the Wiki parser. The
actual code is coloured based on information from the SWI-Prolog cross
referencer using code shared with
PceEmacs\footnote{\url{http://www.swi-prolog.org/emacs.html}}. The
colouring engine uses \index{read_term/3}\predref{read_term}{3} with options `subterm_positions'
to get the term layout compatible to Quintus Prolog
\cite{QUINTUS:manual} and `comments' to get comments and their positions
in the source file.

\postscriptfig[width=\linewidth]{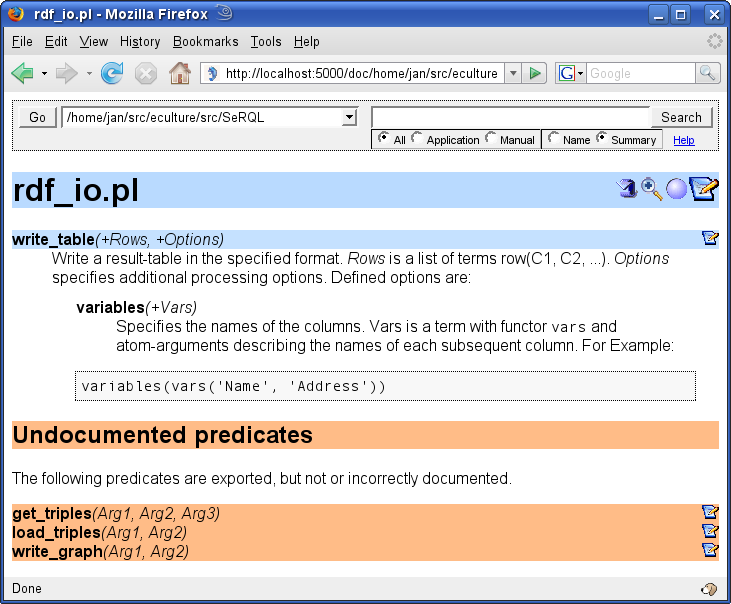}{Undocumented public predicates
are added at the bottom. When accessed from `localhost', the developer
can click the \textit{edit} icon, add or fix documentation and click the
\textit{reload} icon at the top of the page to view the updated
documentation.}

\section{User experiences}
\label{sec:toko}

tOKo \cite{tOKo} is an open source tool for text analysis, ontology
development and social science research (e.g.\ analysis of Web 2.0
documents). tOKo is written in SWI-Prolog. The user base is very diverse
and ranges from semantic web researchers who need direct access to the
underlying code for their experiments, system developers who use an HTTP
interface to integrate a specific set of tOKo functionality into their
systems, to social scientists who only use the interactive user
interface.

The source code of tOKo, 135,000 lines (excluding dictionaries)
distributed over 321 modules provides access to dictionaries, the
internal representation of the text corpus, natural language processing
and statistical NLP algorithms, (pattern) search algorithms, conversion
predicates and the
XPCE\footnote{\url{http://www.swi-prolog.org/packages/xpce/}} code for
the user interface.

Before the introduction of the PlDoc package only part of the user
interface was documented on the tOKo homepage. Researchers and system
developers who needed access to the predicates had to rely on the source
code proper which, given the sheer size, is far from trivial. In
practice, most researchers simply contacted the development team to get
a handle on ``where to start''. This example shows that when open source
software has non-trivial or very large interfaces it is necessary to
complement the source code with proper documentation of at least the
primary API predicates.

After the introduction of PlDoc all new tOKo functionality is being
documented using the PlDoc style of literate programming. The main
advantages have already been mentioned, in particular the immediate
reward for the programmer. The intuitive notation of PlDoc also makes it
relatively easy to add the documentation.
The Emacs Prolog mode developed for
SWI-Prolog\footnote{\url{http://turing.ubishops.ca/home/bruda/emacs-prolog}}
automatically reformats the documentation, such that mixing code and
documentation becomes natural after a very short learning curve.

One of the biggest advantages of writing documentation at all is that it
reinforces a programmer to think about the names and arguments of
predicates. For many of the predicates in tOKo the form is
\term{operation}{Output, Options} or \term{operation}{Input, Output,
Options}. Using an option list, also common in the ISO standard
predicates and the SWI-Prolog libraries, avoids an explosion of
predicates. For example, \index{misspellings_corpus/2}\predref{misspellings_corpus}{2}, which finds misspellings
in a corpus of documents, has options for the algorithm to use, the
minimum word length and so forth: \term{misspellings_corpus}{Output,
[minimum_length(5), edit_distance(1), dictionary(known)]}. Without
documentation, once the right predicate is found, the programmer still
has to check and understand the source code to determine which options
are to be used. Writing documentation forces the developer to think
about determining a consistent set of names of predicates and names of
option type arguments.

A problem that the PlDoc approach only solves indirectly is when complex
data types are used. In tOKo this for example happens for the
representation of the corpus as a list of tokens. In a predicate one can
state that its first argument is a list of tokens, but a list of
tokens itself has no predicate and the documentation of what a token
list looks like is non-trivial to create a link to.  Partial solutions
are to point to a predicate where the type is defined, possibly from a
\index{@see}\objectname{see} keyword or point to a \fileext{txt} file where the type is defined.

Completing the PlDoc style documentation for tOKo is still a daunting
task. The benefits for the developer are, however, too attractive not to
do it.

\section{Related work}
\label{sec:related}

The lpdoc system \cite{DBLP:conf/cl/Hermenegildo00} is the most widely
known literate programming system in the Logic Programming world. It
uses a rich annotation format represented as Prolog directives and
converts these into Texinfo \cite{texinfo}. Texinfo has a long history,
but in our view it is less equipped for supporting interactive literate
programming for Logic Programming in a portable environment. The
language lacks the primitives and notational conventions in the Logic
Programming domain and is not easily expanded. The required \TeX{} based
infrastructure and way of thinking no longer is a given.

In Logtalk \cite{pmoura03}, documentation supporting declarations are
part of the language. The intermediate format is XML, relying on XML
translation tools and style sheets for rendering in browsers and on
paper.  At the same time the structure information embedded in the
XML can be used by other tools to reason about Logtalk programs.

The
ECLiPSe\footnote{\url{http://eclipse.crosscoreop.com/doc/userman/umsroot088.html}}
documentation tools use a single \index{comment/1}\predref{comment}{1} directive containing an
attribute-value list of information for the documentation system. The
Prolog based tools render this as HTML or plain text.

PrologDoc\footnote{\url{http://prologdoc.sourceforge.net/}} is a Prolog
version of JavaDoc. It stays close to JavaDoc, heavily relying on
`@'-keywords and using HTML for additional markup.  \Figref{prologdoc}
gives an example.

\begin{figure}
\begin{verbatim}
/**
    @form member(Value,List)
    @constraints
    @ground Value
    @unrestricted List
    @constraints
        @unrestricted Value
        @ground List
    @descr True if Value is a member of List
*/
\end{verbatim}

\noindent
    \caption{An example using PrologDoc}
    \label{fig:prologdoc}
\end{figure}

Outside the Logic Programming domain there is a large set of literate
programming tools.  A good example, the website of which contains a lot
of information on related systems, is Doxygen \cite{doxygen}.  Most of
the referenced systems use structured comments.

\section{Extending and porting PlDoc}
\label{sec:future}

Although to us the embedded HTTP server backend is the primary target,
PlDoc will be extended with backends for static HTML files (partially
realised). PlDoc is primarily an API documentation system. It is
currently not very suitable for generating a book. Such functionality is
highly desirable for dealing with the SWI-Prolog system documentation,
maintained in \LaTeX{}. We will investigate the possibility to introduce
a \LaTeX{} macro that will extract the documentation of a file or single
predicate and insert it into the \LaTeX{} text. For example:

\begin{verbatim}
    \begin{description}
        \pldoc{member}{2}
        \pldoc{length}{2}
    \end{description}
\end{verbatim}

\noindent
PlDoc is Open Source and can be used as the basis for other Prolog
implementations. The required comment processing hooks can be
implemented easily in any Prolog system. The comment gathering and
processing code requires a Quintus style module system. The current
implementation uses SWI-Prolog's nonstandard (but not uncommon) packed
string datatype for representing comments.  Avoiding packed strings is
possible, but increases memory usage on most systems.

The web server relies on the SWI-Prolog HTTP package, which in turn
relies on the socket library and multi-threading support. Given the
standardisation effort on thread support in
Prolog\footnote{\url{http://www.sju.edu/~jhodgson/wg17/projects.html}},
portability may become feasible. In many situations it may be acceptable
and feasible to use the SWI-Prolog hosted PlDoc system while actual
development is done in another Prolog implementation.

\section{Conclusions}

In literate programming systems there are choices on the integration
between documentation and language, the language used for the
documentation and the backend format(s). Getting programmers to document
their code is already hard enough, which provided us with the motivation
to go for minimal work and maximal and immediate reward for the
programmer. PlDoc uses structured comments using Wiki-style
documentation syntax extended with plain-text conventions from the
Prolog community. The primary backend is HTML+CSS, served from an HTTP
server embedded in Prolog. The web application provides a unified search
and view for the application code, Prolog libraries and Prolog reference
manual.

\subsection*{Acknowledgements}

Although the development of PlDoc was on our radar for a long time,
financial help from a commercial user in the UK finally made it happen.
Comments from the SWI-Prolog user community have helped fixing bugs and
identifying omissions in the functionality.

\bibliographystyle{plain}
\bibliography{pl,pldoc}

\begin{thebibliography}{10}

\bibitem{QUINTUS:manual}
AI International ltd., Berkhamsted, UK.
\newblock {\em Quintus Prolog, User Guide and Reference Manual}, 1997.

\bibitem{tOKo}
Anjo Anjewierden, Bob Wielinga, and Robert de~Hoog.
\newblock Task and domain ontologies for knowledge mapping in operational
  processes.
\newblock Metis Deliverable 4.2/2003, University of Amsterdam, 2004.
\newblock tOKo home: http://www.toko-sigmund.org/.

\bibitem{texinfo}
Robert~J. Chassell and Richard~M. Stallman.
\newblock {\em {Texinfo}: The {GNU} Documentation Format}.
\newblock Reiters.com, 1999.

\bibitem{Deransart:96}
P.~Deransart, A.~Ed-Dbali, and L.~Cervoni.
\newblock {\em Prolog: The Standard}.
\newblock Springer-Verlag, New York, 1996.

\bibitem{DBLP:conf/cl/Hermenegildo00}
Manuel~V. Hermenegildo.
\newblock A documentation generator for (c)lp systems.
\newblock In John~W. Lloyd, Ver{\'o}nica Dahl, Ulrich Furbach, Manfred Kerber,
  Kung-Kiu Lau, Catuscia Palamidessi, Lu\'{\i}s~Moniz Pereira, Yehoshua Sagiv,
  and Peter~J. Stuckey, editors, {\em Computational Logic}, volume 1861 of {\em
  Lecture Notes in Computer Science}, pages 1345--1361. Springer, 2000.

\bibitem{DBLP:conf/acsc/JefferyHS00}
David Jeffery, Fergus Henderson, and Zoltan Somogyi.
\newblock Type classes in mercury.
\newblock In {\em ACSC}, pages 128--135. IEEE Computer Society, 2000.

\bibitem{479}
Donald~E. Knuth.
\newblock Literate programming.
\newblock {\em Comput. J.}, 27(2):97--111, 1984.

\bibitem{wikiway}
B.~Leuf and W.~Cunningham.
\newblock {\em The Wiki Way: Collaboration and Sharing on the Internet}.
\newblock Addison-Wesley, 2001.

\bibitem{pmoura03}
Paulo Moura.
\newblock {\em {Logtalk - Design of an Object-Oriented Logic Programming
  Language}}.
\newblock PhD thesis, Department of Informatics, University of Beira Interior,
  Portugal, September 2003.

\bibitem{ajax}
Linda~Dailey Paulson.
\newblock {B}uilding {R}ich {W}eb {A}pplications with {A}jax.
\newblock {\em IEEE Computer}, 38(10):14--17, 2005.

\bibitem{1035054}
Vreda Pieterse, Derrick~G. Kourie, and Andrew Boake.
\newblock A case for contemporary literate programming.
\newblock In {\em SAICSIT '04: Proceedings of the 2004 annual research
  conference of the South African institute of computer scientists and
  information technologists on IT research in developing countries}, pages
  2--9, , Republic of South Africa, 2004. South African Institute for Computer
  Scientists and Information Technologists.

\bibitem{ramsey91literate}
Norman Ramsey and Carla Marceau.
\newblock Literate programming on a team project.
\newblock {\em Software - Practice and Experience}, 21(7):677--683, 1991.

\bibitem{AOPS}
A.~Shum and C.~Cook.
\newblock Aops: an abstraction-oriented programming system for
  literateprogramming.
\newblock {\em Software Engineering Journal}, 8(3):113--120, 1993.

\bibitem{191059}
Stephen Shum and Curtis Cook.
\newblock Using literate programming to teach good programming practices.
\newblock In {\em SIGCSE '94: Proceedings of the twenty-fifth SIGCSE symposium
  on Computer science education}, pages 66--70, New York, NY, USA, 1994. ACM
  Press.

\bibitem{806466}
Richard~M. Stallman.
\newblock Emacs the extensible, customizable self-documenting display editor.
\newblock {\em SIGPLAN Not.}, 16(6):147--156, 1981.

\bibitem{doxygen}
D~van Heesch.
\newblock {\em Doxygen, a documentation system for C++}, 2007.
\newblock http://www.stack.nl/~dimitri/doxygen/.

\bibitem{TPLP06}
Jan Wielemaker, Zhisheng Huang, and Lourens van~der Mey.
\newblock {SWI-Prolog} and the {Web}.
\newblock Accepted for publication in tplp, HCS, University of Amsterdam, 2006.

\end{thebibliography}

\egroup

\end{document}